\documentclass[epsf,usegraphicx]{mn2e}

\title[A search for electron cyclotron maser emission from compact
binaries] {A search for electron cyclotron maser emission from compact
binaries}

\author[]
{Gavin Ramsay$^{1,2}$, Catherine Brocksopp$^{2}$, Kinwah Wu$^{2}$, Bruce 
Slee$^{3}$, Curtis J. Saxton$^{2}$\\
$^{1}$Armagh Observatory, College Hill, Armagh, BT61 9DG, Northern Ireland, 
UK\\ 
$^{2}$Mullard Space Science Lab, University College London,
Holmbury St. Mary, Dorking, Surrey, RH5 6NT, England, UK\\
$^{3}$Australia Telescope National Facility, CSIRO, PO Box 76, Epping, 
NSW 1710, Australia\\
}

\begin{document}
\outer\def\gtae {$\buildrel {\lower3pt\hbox{$>$}} \over 
{\lower2pt\hbox{$\sim$}} $}
\outer\def\ltae {$\buildrel {\lower3pt\hbox{$<$}} \over 
{\lower2pt\hbox{$\sim$}} $}
\newcommand{\ergscm} {ergs s$^{-1}$ cm$^{-2}$}
\newcommand{\ergss} {ergs s$^{-1}$}
\newcommand{\ergsd} {ergs s$^{-1}$ $d^{2}_{100}$}
\newcommand{\pcmsq} {cm$^{-2}$}
\newcommand{\ros} {\sl ROSAT}
\newcommand{\chan} {\sl Chandra}
\newcommand{\xmm} {\sl XMM-Newton}
\def\rchi{{${\chi}_{\nu}^{2}$}}
\newcommand{\Msun} {$M_{\odot}$}
\newcommand{\Mwd} {$M_{wd}$}
\def\Mdot{\hbox{$\dot M$}}
\def\mdot{\hbox{$\dot m$}}
\newcommand{\teff}{\ensuremath{T_{\mathrm{eff}}}\xspace}
\newcommand{\ratus}{RAT\,J0455+1305\xspace}

\maketitle

\begin{abstract}

Unipolar induction (UI) is a fundamental physical process, which
occurs when a conducting body transverses a magnetic field. It has
been suggested that UI is operating in RX~J0806+15 and RX~J1914+24,
which are believed to be ultra-compact binaries with orbital periods
of 5.4~min and 9.6~min respectively. The UI model predicts that those
two sources may be electron cyclotron maser sources at radio
wavelengths. Other systems in which UI has been predicted to occur are
short period extra-solar terrestrial planets with conducting cores. If
UI is present, circularly polarised radio emission is predicted to be
emitted.  We have searched for this predicted radio emission from
short period binaries using the VLA and ATCA. In one epoch we find
evidence for a radio source, coincident in position with the optical
position of RX~J0806+15. Although we cannot completely exclude that
this is a chance alignment between the position of RX~J0806+15 and an
artifact in the data reduction process, the fact that it was detected
at a significance level of 5.8$\sigma$ and found to be transient,
suggests that it is more likely that RX~J0806+15 is a transient radio
source. We find an upper limit on the degree of circular polarisation
to be $\sim50\%$.  The inferred brightness temperature exceeds
$10^{18}$~K, which is too high for any known incoherent process, but
is consistent with maser emission and UI being the driving
mechanism. We did not detect radio emission from ES Cet, RX~J1914+24
or Gliese 876.

\end{abstract}


\section{Introduction}

Unipolar induction (UI) is a fundamental electromagnetic process.  For
astrophysical systems containing a magnetic and a non-magnetic body
orbiting around a common center of gravity, a large e.m.f.  is induced
across the system by UI when the rotation period of each respective
body deviates from one another or from the binary orbital period. If a
magnetized plasma is present in the binary environment, an electric
current circuit will be set up. The dissipation of the electric
currents will heat the magnetic object, which may cause an
observational signature.  The location where the dissipation occurs
depends on the conductivity of the two objects and the nature of the
plasma between them; it also depends on the magnetic-field
configuration of the system.  If the electrical conductivity of the
two objects is similar, a dipolar magnetic field will lead to strong
heating in regions near the field foot-points of the magnetic object,
where the current density is the highest, as the electric currents are
focused by the converging magnetic field lines.

Among all astrophysical UI systems, the best known system may be the
Jovian system (Piddington \& Drake 1968; Goldreich \& Lynden-Bell
1969).  The currents flowing between the Galilean moons and Jupiter
cause heating on Jupiter, resulting in hot spots and trails on
Jupiter's atmosphere (Connerney et al. 1993; Clarke et al. 1996;
2002). In the Jovian system, the large-scale magnetic field is
provided by Jupiter, while volcanism on the moon Io may supply the
plasma for the conduction of electric currents (Brown 1994). More
recently, two very late-type stars have been found to emit radio
emission which varied in its intensity on a period of a few hours, and
is polarised and highly variable (Berger et al 2005, Hallinan et al.
2006, 2007). Such properties are consistent with those predicted by
the UI model.

It has been suggested that UI processes resembling that found in the
Jovian system can occur in double degenerate binary systems (Wu et
al. 2002), and in degenerate star-planet systems (Li, Ferrario \&
Wickramasinghe 1998; Willes \& Wu 2004, 2005). Two candidates for UI
double degenerate binary systems are the peculiar X-ray sources
RX~J0806+15 and RX~J1914+24. These objects show light curves which are
modulated on periods of 5.4 and 9.5 min respectively - these periods
have widely been taken to represent the binary orbital periods (Ramsay
et al 2000, Ramsay, Hakala \& Cropper 2002, Israel et al 2002). Most
of the models which have been put forward to account for these systems
have two white dwarfs orbiting around a common center of gravity (eg
Cropper et al 2004). Wu et al (2002) proposed that the X-ray emission
from both these systems were powered by UI.  Since their orbital
periods are much shorter than the rotation period of Jupiter's
satellites, and since the magnetic field of the magnetic white dwarf
is much greater than Jupiter, the currents are much greater and
therefore heat the footpoints to X-ray rather than UV temperatures (in
the case of Jupiter).

An essence of the UI model for these systems is the large current
flows, driven by the induced e.m.f., along magnetic field lines
connecting the two objects.  For a magnetic object with a dipolar
field, the field lines converge to its magnetic polar regions.  In
such a field configuration, kinetic instabilities such as loss-cone
instability can develop easily, leading to electron-cyclotron masers
(Wu \& Lee 1979; see also Dulk 1985).  Moreover, while there is
sufficient plasma to provide the charge particles for the current, the
plasma density should be low enough to have a low plasma cut-off
frequency for the transmission of masers.  For double degenerate
binaries and stellar planetary systems, some harmonics of these masers
are in radio wave-bands (Willes \& Wu 2004; Willes, Wu \& Kuncic
2005).  Moreover, these electron-cyclotron masers are narrowly beamed,
have a high brightness temperature ($\gg 10^8$~K) and are almost 100\%
circularly polarized.  Thus, the detection of strongly circularly
polarized emission, (modulated at the orbital period), would provide
an unambiguous proof of the presence of UI in these compact systems.
   
\section{Candidate UI targets}

We sought evidence of UI in various astrophysical systems. These
include the systems RX~J0806+15 and RX~J1914+24 and as a control,
we also observed the ultra-compact system ES~Cet which has an orbital
period of 10.3 min (Warner \& Woudt 2002) and shows clear evidence of
accretion (Espaillat et al. 2005). While the presence of an accretion
flow should suppress UI, it is interesting to search for radio
emission around such short period binaries.

Our other target was Gliese 876 (GJ 876) which is a dM4 star at a
distance of 4.7 pc with 3 known planets. GJ 876d has an orbital period
of 1.94 days and has a mass of $\sim7.5_{\earth}$ (Rivera et al
2005). If the planet GJ 876d has a negligible magnetic field and a
metallic core, UI could occur which may drive electron cyclotron maser
emission, analogous to the Jovian system. On the other hand if GJ 876d
has a substantial surface magnetic field, its field will interact
directly with the field of the host star and shield its core from
being threaded by the field lines of its dM4 host-star. In this case,
field reconnection may lead to particle acceleration, and any radio
emission would probably be due to synchrotron radiation and not
electron cyclotron maser emission. While the radiation would be
expected to be linearly polarized, we would not expect a clear
periodicity related to its orbital motion, unless the emission region
is localised and eclipsed for a fraction of the orbit.

Willes, Wu \& Kuncic (2004) presented calculations for the peak radio
flux expected from ultra-compact binaries such as RX~J0806+15 and 
RX~J1914+24. For systems with orbital periods in the range 5--10 min,
the optimum observing frequency is close to 5 GHz (6cm). For
terrestrial planets orbiting around a low-mass magnetic star,
the UI model described in Willes \& Wu (2005) predicted 
that the peak frequency is likely to be between 50--500 GHz
(0.6--6mm) although this was rather uncertain.

\section{Observations and Results}

ES Cet and GJ 876 were observed using the Australian Telescope Compact
Array (ATCA) in New South Wales, Australia. RX~J0806+15 and
RX~J1914+24 were observed using the Very Large Array (VLA) in New
Mexico, USA. We obtained full polarisation information, with the
intention of determining the fractional polarisation of any detected
source. The observation log is shown in Table \ref{log}. In each case,
observations took place at two adjacent frequency bands; this provides
an increase in sensitivity by a factor of $\sqrt 2$ when imaged
altogether.

All data were reduced using standard flagging, calibration and imaging
routines within the {\sc miriad} and {\sc aips} packages for the ATCA
and VLA observations respectively.

We determined the rms noise level in regions of sky near the known
position of each source. We set upper limits on the flux density to be
three times the rms noise level.  For the one source from which we
detected radio emission (RX J0806+15; see Section 3.3), the integrated
flux density was obtained by performing a 2-d Gaussian fit to the
image.

\begin{table}
\begin{center}
\begin{tabular}{lrrrr}
\hline
Source        & Telescope & $\lambda$ & Date & Duration\\
\hline
ES Cet & ATCA & 6.1 cm & 27 Mar 2005 & 9.3 hrs\\
RX J1914+24   & VLA & 6.2 cm & 12 Sept 2005 & 3 hrs \\
RX J0806+15   & VLA & 6.2 cm & 26 Sept 2005 & 3 hrs \\
RX J0806+15   & VLA & 6.2 cm & 29 Dec 2006  & 10 hrs \\
GJ 876    & ATCA & 12mm & 27 Feb 2006 & 10 hrs \\
GJ 876    & ATCA & 12mm & 28 Feb 2006 & 10 hrs \\
GJ 876    & ATCA & 12mm & 18 Mar 2006 & 10 hrs \\
GJ 876    & ATCA & 12mm & 19 Mar 2006 & 10 hrs \\
\hline
\end{tabular}
\end{center}
\caption{The log for the radio observations presented here. The
Australian Telescope Compact Array is located in Narrabri, Australia,
while the Very Large Array is located in New Mexico, USA.}
\label{log}
\end{table}

\subsection{ATCA observations of ES Cet}

The array was in the 6A configuration and observations took place at
frequency bands centered on 4800 and 4928 MHz. Conditions were good
throughout the observations. PKS~1934$-$638 and PKS~020$-$170 were used as
flux and phase calibration sources respectively. In order to reduce any
potential contamination of a weak source by artifacts at the centre of the
field, the beam was offset by $\sim0.5^{'}$ from the desired
target. No radio source was found at the position of ES Cet and we determine
a $3\sigma$ upper limit to the flux density of 72$\mu$Jy.

\subsection{VLA observations of RX J1914+24}

The array was in the C configuration and the observations took place at
frequency bands centered on 4835 and 4885 MHz. Some scattered cloud was
present. Flux calibration was performed with respect to 3C286 and 3C48; PKS
J1925+2106 was used as the phase reference source. No radio source was found
at the position of RX J1914+24 and we determine a $3\sigma$ upper limit to
the flux density of 42$\mu$Jy.

\subsection{VLA observations of RX J0806+15}

Observations were obtained for this object at two distinct epochs.  In
both cases, the array was in the C configuration and the observations
took place at frequency bands centered on 4835 and 4885 MHz. Flux
calibration was performed with respect to 3C147 for the first epoch
and 3C147 plus 3C286 for the second. BWE 0759+1818 and BWE 0748+1239
were used as phase reference sources for the first and second epochs
respectively.

During the first epoch we detected an unresolved point-like radio
source, coincident, to within the uncertainties, with the optical
position of RX J0806+15 (Ramsay, Hakala \& Cropper 2002). A Gaussian
fit to the image yields an integrated flux density of
$99\pm17\,\mu$Jy, a $5.8\sigma$ detection with a FWHM of
4.9$\times$4.2 arcsec.

The primary calibrator, 3C147, inadvertently used for this observation
was not a polarization calibrator. Instead we were obliged to use
observations of 3C48, obtained two weeks previously for calibration of
RX J1914+24. Whilst not ideal, this allowed us to place some
constraints on the circular polarisation. Otherwise, polarization
calibration was standard and we obtained a $3\sigma$ upper limit to
the circularly polarized flux of 52$\mu$Jy.  This is the equivalent of
an upper limit of $\sim50\%$ fractional polarization.

The second -- and longer -- observation was made 15 months after the
first, and was made to confirm the presence of the radio source
detected in the first observation.  This second epoch observation
resulted in a non-detection, with a $3\sigma$ upper limit to the flux
density of 36$\mu$Jy.  We plot the resultant images in Figure 1, the
left-hand panel showing the first epoch with the apparent source.

We note that the brighter source towards the bottom-left of the field
(Figure 1) varies in its flux between the two epochs by $\sim70
\mu$Jy.  However, an additional source to the right of the region
displayed in Figure 1 remains constant between the two
epochs. Therefore we believe that this variability is real and not a
calibration artifact.

These results suggest that either the first detection was a chance
coincidence with an artifact in the reduction process, or that the
radio source is transient. To investigate this further, we split the
first observation into eight segments of equal duration and made an
image using each individual segment. In only the first, $\sim$ 20 min,
segment was the source detected; using this segment a Gaussian fit
yielded an integrated flux density of $128\pm39\,\mu$Jy.

We cannot completely exclude that the radio detection was a chance
coincidence between the known source position and an artifact in the
data reduction process. However, since the significance is
5.8$\sigma$, and that the source is variable in the first epoch
observation, we believe that it is more likely that we have detected
variable radio emission from RX~J0806+15.  We discuss the implications
of this result in Section 4.

\subsection{ATCA observations of GJ 876}

The observations took place at frequency bands centered on 18448 and
19472 MHz over four different epochs. The weather was reasonably
favourable for 12-mm observations, although cloud during the
afternoons degraded the data quality to some extent. The flux
calibration sources were PKS~1934$-$638 and/or PKS~1921$-$293; phase
referencing took place with respect to QSO B2243-123.

Since the orbital period of GJ 876d is very close to 2 days (1.94
days) it is difficult to obtain ground-based observations which cover
the whole orbit from one site. In order to do this we obtained two
10-hour observations within two days and repeated this eighteen days
later (see Table 1). We determined the orbital phase of each
observation using the ephemeris of Rivera (2005) for GJ 876d (assuming
$i=90^{\circ}$), which defined $\phi=0.0$ as the transit epoch and has
uncertainty 0.03 cycles. In Table 3 we show the phase coverage of our
observations; we obtained coverage for 90\% of the orbital
period. However, we did not detect radio emission at any of these
epochs, individually or combined into a single image. The $3\sigma$
upper limit to the flux density in the combined image was 122$\mu$Jy.

\begin{table}
\begin{center}
\begin{tabular}{lr}
\hline
Source        & Flux Density\\
\hline
ES Cet          & $<72\mu$Jy \\     
RX J1914+24     & $<42\mu$Jy  \\     
RX J0806+15 (1) & 99$\pm17\mu$Jy \\
RX J0806+15 (2) & $<36\mu$Jy \\
GJ 876 & $<122\mu$Jy \\
\hline
\end{tabular}
\end{center}
\caption{The flux density ($\pm1\sigma$ rms noise level) or 
$3\sigma$ upper limits for the sources in our survey.
The upper limit on the 
radio emission from GJ 876 is determined from all four datasets combined.}
\label{results}
\end{table}

\begin{table}
\begin{center}
\begin{tabular}{lr}
\hline
Date of         & Phase\\
Observation     &       \\
\hline
27 Feb 2006 & 0.03--0.24 \\
28 Feb 2006 & 0.55--0.76 \\
18 Mar 2006 & 0.53--0.01 \\
19 Mar 2006 & 0.30--0.51 \\
\hline
\end{tabular}
\end{center}
\caption{The orbital phase of GJ 876d at the time of our observations. We 
used the ephemeris of Rivera (2005) which assumed $i=90^{\circ}$ and 
$\phi$=0.0 was the phase which would result in a transit.}
\label{phase}
\end{table}

\begin{figure*}
\begin{center}
\setlength{\unitlength}{1cm}
\begin{picture}(14,10)
\put(-2.5,0){\includegraphics{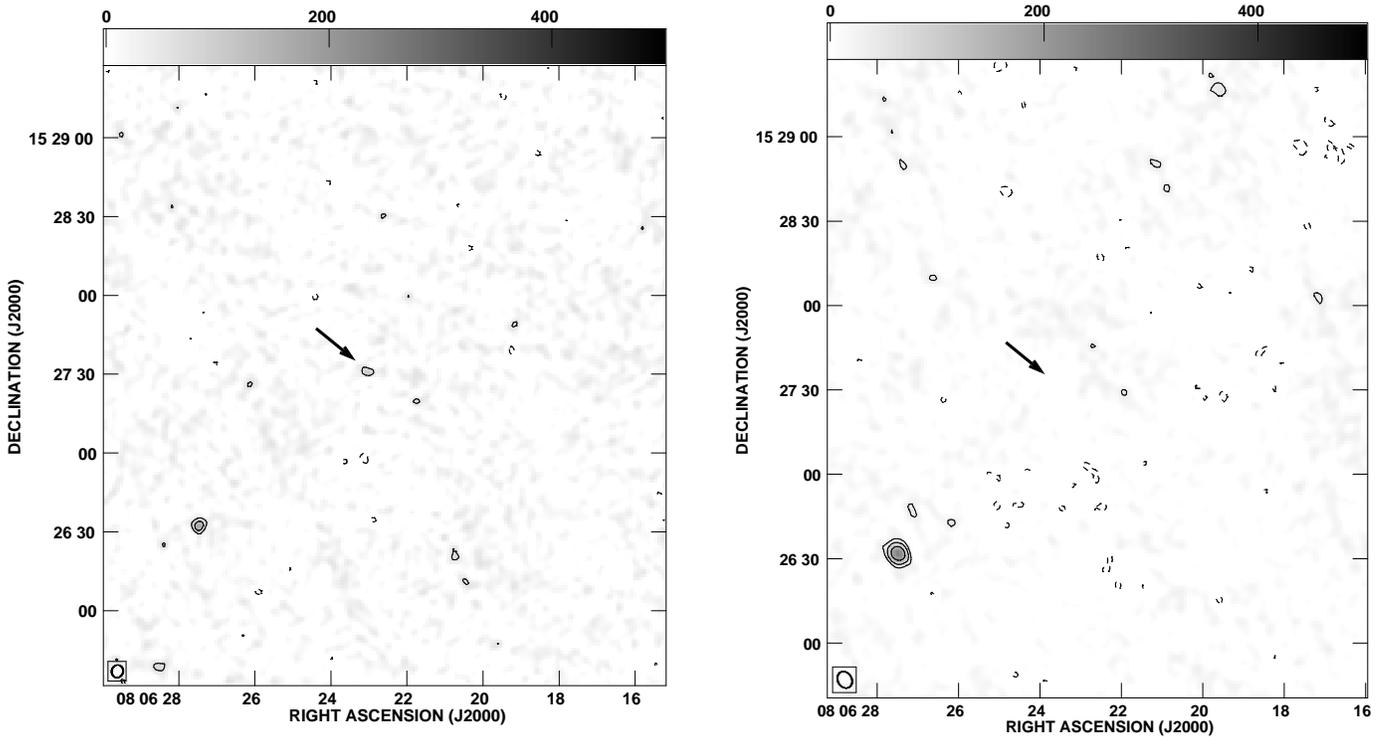}}
\end{picture}
\end{center}
\caption{The radio (6 cm) maps of the field of RX J0806+15 made using
the VLA in Sept 2005 (left hand panel) and in Dec 2006 (right hand
panel). The arrow points to the optical position of RX J0806+15. In
the left hand panel the position of the radio source is 0.3$^{''}$
distant from the optical position of RX J0806+15.  The beam size is
shown in the lower left hand corner of both panels. The rms background
noise (3$\sigma$) is 51$\mu$Jy for the left hand image and 36$\mu$Jy
for the right hand image. The contours are -3, 3, 6, 12, 24, 48, 96
times the 1 sigma level.}
\label{maps}
\end{figure*}

\section{Discussion}

Although UI is believed to operate in various astrophysical systems,
it has still to be determined if UI is an efficient process in double
degenerate binary systems and in magnetised stellar planetary systems.
For double degenerate binaries there is an important issue regarding
the consequences of UI operation, as it may affect the system's energy
budget, and hence the orbital dynamics and evolution (Wu et al. 2002;
Dall'Osso, Israel \& Stella 2006). Moreover, it will also alter the
properties of the gravitational wave emission from these binaries,
which are expected to be detected in large number by the proposed {\sl
LISA} gravitational observatory (see e.g. Nelemans, Yungelson \&
Portegies Zwart 2006).  It is therefore important to verify and assess
the role of UI in compact magnetic binary systems.

The nature of the dominant emission process driving the
electromagnetic radiation from RX~J0806+15 and RX~J1914+24 is not
fully known. While the UI model has successfully accounted for many of
the observational properties of these two systems (eg Dall'Osso,
Israel \& Stella 2007), there are a number of properties which will
require modification of the generic UI model, which assumes a dipole
magnetic field. For instance, it is questionable whether a dipole
field can reproduce, in detail, the phase-offset between the X-ray and
optical data (Barros et al. 2007).  Further, a higher-order magnetic
field component needs to be present in order to produce the X-ray and
optical light curve profile (Barros et al. 2005). Therefore, we need
more direct, alternative evidence to verify the role of UI in contrast
to other processes, such as accretion, in these systems.

In this work we have searched for radio emission from two sources
which are believed to be the two most compact binaries yet known.  If
a magnetic field is present, the electro-magnetic interaction between
the two stars is expected to be stronger than found in magnetic
cataclysmic variables, RS~CVn or magnetic Algols, because of the small
separation between the stars and the rapid orbital rotation (eg
Chanmugam \& Dulk 1982, Retter, Richards \& Wu 2005).  The detection
of radio emission, regardless of its coherence or incoherence, will
provide strong evidence of magnetic interaction, and the detection of
high brightness circularly polarized emission would confirm the
operation of UI on a global scale.

Our observations of ES~Cet showed a null detection.  The lack of
detectable radio emission is not particularly surprising.  ES~Cet is
an ultra-compact binary in which mass transfer occurs via an accretion
flow. The presence of an accretion flow also inhibits any large scale
current circuit, thus, a global UI process as described in Wu et
al. (2002) cannot arise.  Although one cannot exclude UI operating in
a very small local scale, its effects on the orbital dynamics and on
other observational characteristics are not expected to be significant
as in other ultra-compact binaries with accretion flows.  As loss-cone
or other kinetic instabilities cannot develop in accreting systems,
ES~Cet is not expected to be a maser source.  The presence of
high-density material would imply a high plasma-cutoff frequency.  For
a plasma with electron number density of $\sim 5 \times
10^{11}$cm$^{-3}$, the plasma frequency will be well above 5~GHz, thus
it will prevent the propagation of 6 cm radio emission, which is the
observational band of our ATCA observation.

Our observations do not show evidence of radio emission from
RX~J1914+24 at a limit of 42~$\mu$Jy.  One obvious possibility is that
UI may not occur in this system. However, the non-detection does not
rule UI out. As pointed out in Willes \& Wu (2004) the observability
of electron-cyclotron masers from a UI double-degenerate compact
binary depends on the magnetic moment of the magnetic white dwarf
(which determines the frequencies of the cyclotron harmonics), the
amount of thermal electrons filling the electric-current flowing
magnetic flux tubes, the temperature of these thermal electrons, and
the viewing orientation of the binary.  Calculations showed that the
radio emission is detectable in a restrictive region in the parameter
space of UI double-degenerate compact binaries.  Thus, even if UI is
operating efficiently and electron-cyclotron masers are generated in
all systems in an ensemble, some systems will show detectable
electron-cyclotron masers in the radio wavebands, while a significant
fraction of the systems will show null detection in a radio survey.

We have detected a radio source at a position coincident with the
known optical position of RX~J0806+15. Although we cannot completely
exclude that this is a chance alignment between the known position of
RX~J0806+15 and an artifact in the data reduction process, the fact
that it was detected at a significance level of 5.8$\sigma$ and that
the radio source was variable suggests that it is more likely that
RX~0806+15 is a transient radio source.

With these caveats in mind, we can determine the brightness temperature 
$T_{\rm b}$ of a source by:

\begin{eqnarray} 
    T_{\rm b} & \approx & \frac{4}{\pi}\frac{\lambda^2 S_\nu}{k_{\rm B}} 
          \left(\frac{d}{r}\right)^2 \nonumber \\ 
        & = & 3.3\times 10^{16}   \left(\frac{S_\nu}{1~\mu{\rm Jy}} \right) 
             \left(\frac{\lambda}{7.8~{\rm cm}} \right)^2   
           \nonumber \\ 
       & & \hspace*{0.5cm} \times    
          \left(\frac{d}{500~{\rm pc}} \right)^2 
        \left(\frac{r}{2\times 10^7{\rm cm}} \right)^{-2} ~{\rm K} \ .
\nonumber
\end{eqnarray}

The distance of RX~J0806+15 is not well known, with Israel et al
(2003) noting that the distance to the edge of the Galaxy is 500 pc,
while Barros et al (2007) estimate that its distance is greater than
1.1 kpc implying it is out of the Galactic plane. Assuming a
conservative distance of 500~pc and that the size of the emission
region is $\sim 2 \times 10^7$cm (the linear extension of the
foot-point flux-tube for systems with a non-magnetic white dwarf
companion with mass 0.5 M$_\odot$, see Willes \& Wu 2004), the
observed flux density of 99$\mu$Jy implies $T_{\rm
  b}=2.1\times10^{18}$K. There are some uncertainties about the exact
size of the foot-point emission region. Even if we assume that the
size of the emission region is $10^9$cm, (the radius of a 0.5
M$_\odot$ white dwarf), we still obtain a very high brightness
temperature, $T_{\rm b}=8\times10^{14}$K. Such a high brightness
temperature cannot be explained by non-thermal synchrotron process,
which would be limited to $<10^{10}$K (Dulk \& Marsh 1982).  It also
cannot be explained by any incoherent radiation processes, as they are
limited to $\sim 10^{12}$K by inverse Compton cooling (Kellerman \&
Pauliny-Toth 1969).

Therefore, the radio emission must be generated by a coherent
radiative process, such as an electron-cyclotron maser as predicted by
the unipolar-induction model (Wu et al. 2002).  In the maser model
described in Willes \& Wu (2004), the transient or bursting nature of
the source may be explained by variations in the emission-cone beaming
direction or by the presence of a small amount of non-thermal
electrons whose density fluctuate. To confirm the nature of the radio
source we urge further observations of this source at radio
wavelengths to determine how often it shows radio emission and to 
better constrain the upper limit on the circular polarisation.

The fact that we did not detect any evidence for radio emission from
GJ 876 does not rule out the operation of UI in this system. Our
observations of GJ 876 took place at 12 mm. The calculations of Willes
\& Wu (2005) suggest that any radio emission due to the UI operation  
would more likely be observable at shorter wavelengths. Sensitive
observations at these wavelengths ($<$6 mm) will be possible using
ALMA.

\section{Acknowledgements}

The Australia Telescope is funded by the Commonwealth of Australia for
operation as a National Facility managed by the CSIRO. The National
Radio Astronomy Observatory is a facility of the National Science
Foundation of the USA, operated under cooperative agreement by Associated
Universities, Inc.  We thank the staff of both the AT and VLA for
assistance with the observations and Steven Longmore for help in
reducing the ES Cet data. KW thanks Richard Hunstead for discussions.
Armagh Observatory is grant aided by the
N. Ireland Dept. of Culture, Arts and Leisure.
	
{}

\end{document}